\documentclass[prb,twocolumn]{revtex4}

\usepackage{bm,graphicx}

\begin{document}

\title{Efficient algorithms for rigid body integration using optimized
splitting methods and exact free rotational motion}

\author{Ramses van Zon$^*$, Igor P. Omelyan$^\dagger$, and Jeremy
Schofield$^*$}

\affiliation{$^*$Chemical Physics Theory Group, Department of
Chemistry, University of Toronto, 80 Saint George Street, Toronto,
Ontario M5S 3H6, Canada\\
$^\dagger$Institute for Condensed Matter Physics, 1 Svientsitskii
Street, UA-79011 Lviv, Ukraine}

\date{February 8, 2008}

\maketitle

In this note we present molecular dynamics integration schemes that
combine optimized splitting and gradient methods with exact free
rotational motion for rigid body systems and discuss their relative
merits.  The algorithms analyzed here are based on symplectic,
time-reversible schemes that conserve all relevant constants of the
motion.  It is demonstrated that although the algorithms differ in
their stability due to truncation errors associated with limited
numerical precision, the optimized splitting methods can outperform
the commonly-used velocity Verlet scheme at a level of precision
typical of most simulations in which dynamical quantities are of
interest.  Useful guidelines for choosing the best integration scheme
for a given level of accuracy and stability are provided.

Hamiltonian splitting methods are an established technique to derive
stable and accurate integration schemes in molecular
dynamics.\cite{splitting} The strategy of these methods is to split
the Hamiltonian of the system into parts whose evolution can be solved
exactly. Using the Campbell-Baker-Hausdorff formula\cite{cbh},
splitting algorithms can be presented as products of exactly solvable
propagation steps, involving more factors for higher-order
schemes.\cite{ForestRuth} The resulting algorithms can be optimized by
adjusting the form of the splitting to minimize error
estimates.\cite{Omelyan1}

Recently, second- and fourth-order symplectic integration schemes for
simulations of rigid body motion, based on the exact solution for the
full kinetic (free) propagator, have been proposed.\cite{us1} While
this exact solution involves elliptic functions, elliptic integrals
and theta functions,\cite{Jacobi1849} there exist efficient numerical
routines to compute elliptic functions,\cite{gsl} and the computation
of elliptic integrals and theta functions can be implemented
efficiently\cite{us2} or avoided altogether using a recursive
method.\cite{us1} Employing the exact free rotational motion, the
resulting splitting method leads to demonstrably more accurate
dynamics for systems in which free motion is important.\cite{us1}
Furthermore, using the exact kinetic propagator, any splitting scheme
for integrating the dynamics of point particles can be transferred to
rigid systems.
Here we analyze the combination of the exact
kinetic propagator and optimized splitting and
gradient-like\cite{Omelyan1,Omelyana,Omelyan2} approaches.

For a system of rigid bodies, a phase space point $\Gamma$ is specified by a
center of mass position $\mathbf q_i$, an attitude matrix $\mathsf
S_i$, and translational and angular momenta $\mathbf p_i$ and
$\bm\ell_i$ for each particle $i$ of mass $m_i$.
Given
the Hamiltonian $H=T+V$, where $T$ and $V$ are the kinetic and
potential energies, respectively, the time evolution of the point
$\Gamma$ in phase space is governed by $\dot \Gamma =
\{H,\Gamma\}=\{T,\Gamma\}+\{V,\Gamma\}$, in which $\{,\}$ denotes the
Poisson bracket. Henceforth, the operators $\{T,.\}$ and $\{V,.\}$
will be designated as $A$ and $B$, respectively. Defining 
$\mathcal L$ = $A+B$, the solution of the
equations of motion is formally given by $\Gamma(t)=e^{\mathcal L
t}\Gamma(0)$.

While the various possible splitting schemes can be assigned a
theoretical efficiency,\cite{Omelyan1} the relative efficiency of real
simulations can be somewhat different. Nonetheless, the estimates are
useful to eliminate the least efficient variants.  Based on our
studies of second and fourth order methods, the most efficient
integration schemes can be formulated using the following generic form
of the splitting algorithm for a single time step of size $h$:
\begin{equation}
 e^{\mathcal L h} = e^{\eta B h}e^{Ah/2}e^{(1-2\eta)\hat B(\xi)h}
                    e^{Ah/2}e^{\eta B h}
                    +\mathcal O(h^{k+1}).
\label{generic}
\end{equation}
This propagator is applied $t/h$ times to compute the time evolution
of the system over a time interval $t$.  Here, $\eta$ and $\xi$ are
two real parameters, $k$ is the order of the integration scheme, and
$e^{A h}$ and $e^{B h}$ act on a phase space point $\Gamma=\{\mathbf
q_i,\mathbf p_i,\mathsf S_i,\bm\ell_i\}$ as
\begin{eqnarray}
  e^{A h}\Gamma
  &=& \{\mathbf q_i + h \mathbf p_i/m_i, \mathbf p_i,
        \mathsf{P}_i(h)\mathsf{S}_i, \bm\ell_i\} ,
\\
  e^{B h}\Gamma
  &=& \{\mathbf q_i,\mathbf p_i + h\mathbf{f}_i,
        \mathsf S_i,\bm\ell_i + h \bm \tau_i \}
,
\end{eqnarray}
where $\mathbf{f}_i$ and $\bm \tau_i$ are the instantaneous forces and
torques on body $i$, while the matrix $\mathsf{P}_i(h)$ propagates
exactly $\mathsf{S}_i$ over the time interval $h$ in the absence of
torques [see Ref.~\onlinecite{us1} for specific forms for
$\mathsf{P}_i(h)$].  Finally, $\hat B(\xi)$ in Eq.~(\ref{generic}) is
a variation of $B$ which takes the gradients of forces and torques
into account by an advanced gradient-like method.\cite{Omelyan2} More
precisely, the action of $e^{\hat B(\xi) h}$ on a phase space point is
given by
\begin{equation}
  e^{\hat B(\xi)h}\Gamma
=\{\mathbf q_i,\mathbf p_i  + h\tilde\mathbf{f}_i,\mathsf S_i,\bm\ell_i
+ h \bm{\tilde\tau}_i \},
\end{equation}
where the modified forces $\tilde\mathbf{f}_i$ and torques $\bm
{\tilde\tau}_i$ are\cite{Omelyan2}
\begin{equation}
\tilde\mathbf{f}_i = \mathbf{f}_i + \Delta \mathbf{f}_i (\xi,
\lambda) , \quad
\bm{\tilde\tau}_i = \bm\tau_i + \Delta \bm\tau_i (\xi,
\lambda) .
\end{equation}
The shifts in forces and torques account for commutator corrections
involving gradients.\cite{Omelyan2} To fourth order in $h$, the shifts
can be approximated by a finite difference approach using a small
parameter $\lambda$ according to
\begin{eqnarray}
  \Delta \mathbf{f}_i (\xi,\lambda) &=&
  [
    \mathbf{f}_i(\tilde\mathbf q, \tilde\mathsf{S})
  - \mathbf{f}_i(\mathbf q, \mathsf{S})
  ]/\lambda,
\nonumber \\ [-6pt] \\ [-6pt]
  \Delta \bm\tau_i (\xi,\lambda) &=&
  [
    \bm\tau_i(\tilde\mathbf q, \tilde\mathsf{S})
   - \bm\tau_i(\mathbf q, \mathsf{S})
  ]/\lambda,
\nonumber
\end{eqnarray}
where $\mathbf{f}_i(\tilde\mathbf q, \tilde\mathsf{S})$ and
$\bm\tau_i(\tilde\mathbf q, \tilde\mathsf{S})$ are the forces and
torques at the auxiliary coordinates
\begin{eqnarray}
  \tilde\mathbf q_i
  = \mathbf q_i + 2 \xi\lambda h^2 \mathbf{f}_i/m_i
,\quad
  \tilde\mathsf{S}_i
=\mathsf R(2\xi\lambda h^2\mathsf J^{-1}_i\mathsf S_i\bm\tau_i)\mathsf S_i.
\label{shiftA}
\end{eqnarray}
Here, $\mathsf J_i$ is the diagonalized moment of inertia tensor of
the $i$th body [i.e.\ $\mathrm{diag}(I_1,I_2,I_3)$] and
$\mathsf{R}(\mathbf v)$ is the Rodrigues matrix\cite{Goldstein} that
performs a rotation around a vector $\mathbf v$.  Note that for
$\xi=0$, $\hat B(0)=B$, in which case there are no advanced-gradient
contributions.  Although the finite difference approach introduces
non-symplectic terms of order $\lambda^2h^4$, no
discernible energy drift was found for small integration time
steps $h$ when the value of the parameter $\lambda$ was taken to be
roughly $10^{-4}$.\cite{Omelyan2}

\begin{figure}[t]
\centerline{\includegraphics[width=0.76\columnwidth]{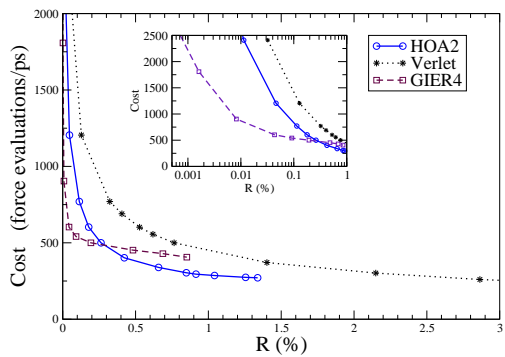}}
\caption{Efficiency of integration schemes for simulations of rigid
water. For various values of the timestep $h$, the plot shows the
relative error versus cost (in force evaluation per ps). The plots
extend up to values of $R$ where the simulations start to exhibit
statistically-significant drift due to numerical round-off. The inset
shows the same on a logarithmic scale.}
\label{fig}
\end{figure}

By tuning the parameters $\xi$ and $\eta$, different integration
schemes can be obtained. Choosing $\xi=0$ and $\eta=0$ or $\eta=1/2$
results in the well-known second-order ($k=2$) Verlet scheme, in its
position or velocity form, respectively. Fixing $\xi=0$ but allowing
$\eta$ to vary, the prefactors can be minimized in front of the
$\mathcal O(h^2)$ corrections, which gives $\eta=0.1931833275037836$
as an optimal choice. \cite{Omelyan1,Omelyan2} This scheme, which was
called HOA2 in Ref.~\onlinecite{Omelyan2}, is still second order but
is expected to be more accurate. Finally, one can vary both $\eta$ and
$\xi$, to make the prefactors of the $\mathcal O(h^2)$ corrections
vanish to yield a fourth-order algorithm. \cite{Omelyan2} For this
scheme, which we have called GIER4, the required values are $\eta=1/6$
and $\xi=1/48$.

To assess the relative computational cost of each of the integration
schemes at a given level of accuracy, simulations of 512 rigid water
molecules using the TIP4P potential\cite{TIP4P} were carried out at
liquid density of 1 g/cm$^3$ and a temperature of 297~K.  The accuracy
of the simulations was measured by calculating the ratio $R$ of
fluctuations of the total energy to the fluctuations of the potential
energy at a given computational load.  This load was estimated by
using the number of force evaluations in a given time interval, here
taken to be 1 picosecond (ps).  At liquid densities, the computational
load correlates very well with the overall CPU time since relatively
little CPU time is required in the free motion propagation steps.  In
addition, the stability of each integration scheme was monitored by a
linear least-squared analysis of the drift of the total energy over a
series of 10 to 50 runs of total length $15$ ps for each time step
reported.

The results of this analysis are plotted in Fig.~1, from which it is
evident that for crude simulations requiring only modest energy
conservation (i.e. $R>1.5 \%$), the standard Verlet algorithm is the
only algorithm that is stable.  Trajectories at this level of accuracy
can be used in sampling schemes such as hybrid Monte-Carlo.  However
for $R < 1.5 \%$, arguably the upper limit of allowable error in
simulations from which dynamical information can be extracted, the
optimized second-order HOA2 scheme is roughly $1.5$ times more
efficient than the Verlet algorithm.  Note that the HOA2 algorithm
differs from the velocity Verlet scheme only in the choice of time
step for the momenta updates and is therefore simple to implement.
Interestingly, the fourth order GIER4 scheme is preferable if very
accurate simulations are required ($R < 0.4 \%$) in spite of the
additional computational cost of the modified forces and torques at
auxiliary positions.  Other fourth-order splitting
schemes\cite{Omelyan1} (not outlined here) have also been tested and
found to be less efficient than the relatively simple GIER4.
Streamlining explicit calculations of the gradients of forces and
torques instead of utilizing finite difference methods would restore
symplecticity and likely increase the value of $R$ at which the GIER4
method is optimal.

\noindent\textbf{Acknowledgments:} R.v.Z. and J.S.  acknowledge
support by a grant from NSERC and a PRF (ACS) grant. I.O. thanks the
Fonds zur F\"orderung der wissenschaftlichen Forschung (project
No. 18592-PHY).


\begin{thebibliography}{99}

\bibitem{splitting}
 B. Leimkuhler and S. Reich,
 \emph{Simulating Hamiltonian Dynamics}
 (Cambridge University Press, Cambridge, 2005).

\bibitem{cbh}
 G. Parisi, 
 \emph{Statistical Field Theory} 
  (Addison-Wesley, Reading, MA, 1988).

\bibitem{ForestRuth}
 E.\ Forest and R.D.\ Ruth, Physica D \textbf{43}, 105 (1990).

\bibitem{Omelyan1}
 I.\ P.\ Omelyan, I.\ M.\ Mryglod, and R.\ Folk,
 Comp.\ Phys.\ Comm.\ \textbf{151}, 272­314 (2003).

\bibitem{us1}
 R.\ van Zon and J.\ Schofield,
 Phys.\ Rev.\ E \textbf{75}, 056701 (2007).

\bibitem{Jacobi1849}
 C.G.J.\ Jacobi,
 Crelle J.\ Reine Angew.\ Math. \textbf{39}, 293 (1849).



\bibitem{gsl}
M.\ Galassi, J.\ Davies, J.\ Theiler, B.\ Gough, G.\ Jungman, M.\
Booth, and F.\ Rossi, \emph{GNU Scientific Library Reference Manual}
(Network Theory Ltd, Bristol, UK, 2005), revised 2nd ed.

\bibitem{us2}
 R.\ van Zon and J.\ Schofield,
 J.\ Comput.\ Phys.\ \textbf{225}, 145 (2007).

\bibitem{Omelyana}
 I.P.\ Omelyan, Phys.\ Rev.\ E \textbf{74}, 036703 (2006).

\bibitem{Omelyan2}
 I.P.\ Omelyan,
 J.\ Chem.\ Phys.\ \textbf{127}, 044102 (2007).

\bibitem{Goldstein}
 H.\ Goldstein, \emph{Classical Mechanics} (Addison-Wesley, Reading,
 Massachusetts, 1980).

\bibitem{TIP4P}
 W.L.\ Jorgensen, J.\ Chandrasekhar, J.D.\ Madura, R.W.\ Impey, and
 M.L.\ Klein, J.\ Chem.\ Phys.\ \textbf{79}, 926 (1983).

\end{thebibliography}
\end{document}